\documentclass[11pt,letter]{article}
\usepackage[margin=1in]{geometry}
\usepackage[toc,page,title,titletoc,header]{appendix}
\usepackage{setspace}

\usepackage{hyperref}
\usepackage{graphicx}
\usepackage{multirow}
\usepackage{booktabs}
\usepackage{threeparttable}
\usepackage[margin=1cm,labelfont=bf]{caption}
\usepackage[noblocks]{authblk}

\usepackage[abbr]{harvard}

\newenvironment{figurenotes}
  {\par
   \flushleft \linespread{1} \small
   \emph{Notes}: \ignorespaces
  }
  {\par}
  
\begin{document}

\title{Facebook Shadow Profiles\thanks{The authors thank Stefan Bechtold, Dirk Bergemann, J\"org Claussen, Tomaso Duso, Lisa George, Avi Goldfarb, Joel Waldfogel, Ken Wilbur, and participants of presentations at the Berlin Centre for Consumer Policies (BCCP) Forum, the Media Economics Workshop Stellenbosch, University of Bremen, ETH Zurich Center for Law and Economics, University of Zurich, HEC Lausanne, and IPTS Seville for valuable feedback. We are grateful to Ilia Azizi for excellent research assistance and to Ulrich Kaiser for enabling access to the proprietary data used in the study. Christian Peukert is grateful for funding by the Swiss National Science Foundation under Grant Number 100013\_197807.}}

  \author[a]{Luis Aguiar}
  \author[b]{Christian Peukert} 
  \author[c,d]{Maximilian Sch\"afer} 
  \author[e,f]{Hannes Ullrich}

  \affil[a]{Department of Business Administration, University of Zurich}
  \affil[b]{Department of Strategy, Globalization and Society, HEC, University of Lausanne}
  \affil[c]{Tobin Center for Economic Policy, Yale University}
  \affil[d]{Department of Economics, University of Bologna}
  \affil[e]{Department of Firms and Markets, DIW Berlin}
  \affil[f]{Department of Economics, University of Copenhagen}

\date{May 2022}

\maketitle

\thispagestyle{empty} \setcounter{page}{0}
\begin{abstract}
\noindent 
We quantify Facebook's ability to build shadow profiles by tracking individuals across the web, irrespective of whether they are users of the social network. For a representative sample of US Internet users, we find that Facebook is able to track about 40\% of the browsing time of both users and non-users of Facebook, including on privacy-sensitive domains and across user demographics. We show that the collected browsing data can produce accurate predictions of personal information that is valuable for advertisers, such as age or gender. Because Facebook users reveal their demographic information to the platform, and because the browsing behavior of users and non-users of Facebook overlaps, users impose a data externality on non-users by allowing Facebook to infer their personal information.

\end{abstract}

\onehalfspacing


\newpage
\section{Introduction}

The fundamental business model of many online platforms consists in generating revenue through online advertising. Because detailed information about consumers' types and preferences is crucial for the effectiveness of targeted advertising, online platforms have developed innovative technologies to collect and analyze behavioral data.

In the particular case of social media, online platforms can naturally rely on the information directly provided by their users to construct consumer profiles based on demographics and personal interests. However, they can also go beyond the direct profiling of their users and construct \emph{shadow profiles} of individuals who are not users of their service \cite{garcia2017shadowprof}. The main mechanism documented in the literature relies on users of a particular platform sharing basic information about non-users, for instance their contact list. While such explicit revelation by users is sufficient for platforms to construct shadow profiles of non-users, we show that it is not necessary, nor is the knowledge of the social network structure connecting users with non-users of the platform. 

More specifically, we show that a large platform like Facebook can meaningfully predict non-users' personal information and consequently build shadow profiles by solely relying on users' personal information revealed to the platform - such as age or gender - coupled with information on users'  and non-users' browsing activity \emph{outside} of the platform. Therefore, when sharing their personal information with Facebook, users of the platform exert a data externality on non-users by enabling the platform to construct shadow profiles without their explicit consent \cite{maccarthy2010new,choi2019privacy,acemoglutoo}. Facebook may thus monetize data on individual browsing behavior by selling targeted ads based on shadow profiles through its cross-site advertising network. In addition, browsing histories may be linked to existing and future Facebook accounts or be used to predict missing data points on Facebook users with similar browsing characteristics.\footnote{See also Facebook's official press release on data collection outside of Facebook at \href{https://about.fb.com/news/2018/04/data-off-facebook/}{https://about.fb.com/news/2018/04/data-off-facebook/}, accessed 24 September 2021.}

The mechanism that allows Facebook to build shadow profiles based on browsing behavior alone relies on its network of engagement buttons. Using cookies, Facebook can track individuals - regardless of whether they are Facebook users - across all websites on which a Facebook ``Like'' or ``Share'' button appears, even if users never actively click on them \cite{roosendaal12}.  A large number of websites have integrated Facebook's engagement buttons since their launch in 2009, perhaps with the intent to increase traffic through social media referrals, and their widespread diffusion on the web has been documented in academic work and industry reports \cite{chaabane2012big,libert2015exposing,lerner2016internet,EnglehardtNarayanan2016,ghostery20}. In 2020, Facebook's engagement buttons had reached 40 percent of the top 1'000 websites \cite{accc2021}. 

The ability of Facebook to build shadow profiles has also raised substantial public interest in recent years.\footnote{See, for example, \href{https://eu.usatoday.com/story/tech/columnist/baig/2018/04/13/how-facebook-can-have-your-data-even-if-youre-not-facebook/512674002/}{https://eu.usatoday.com/story/tech/columnist/baig/2018/04/13/how-facebook-can-have-your-data-even-if-youre-not-facebook/512674002/}, accessed 24 September 2021.} In 2018, when questioned in U.S. congress and European parliament hearings, CEO Mark Zuckerberg asserted that he did not know about shadow profiles nor how much data on non-users of Facebook was collected.\footnote{See \href{https://venturebeat.com/2018/05/22/mark-zuckerberg-dodges-question-from-european-parliament-on-facebook-shadow-profiles/}{https://venturebeat.com/2018/05/22/mark-zuckerberg-dodges-question-from-european-parliament-on-facebook-shadow-profiles/} and \href{https://techcrunch.com/2018/04/11/facebook-shadow-profiles-hearing-lujan-zuckerberg}{https://techcrunch.com/2018/04/11/facebook-shadow-profiles-hearing-lujan-zuckerberg}, accessed 24 September 2021.} At the same time, studies documenting the extent of tracking technologies such as Facebook's mainly focused on the share of websites being monitored, leaving aside the measurement of individual users' share of online activity being tracked. Documenting the magnitude of individuals' online attention being tracked by Facebook, and the role this information can play in the construction of shadow profiles, is therefore of high relevance for both public policy as well as for advertisers.  

Against this backdrop, the goal of this paper is to tackle three main questions. First, what share of an individual's browsing activity can be tracked by Facebook? Second, how do these shares vary between individuals who are users of Facebook and those who are not? Third, can Facebook use the tracked web browsing activity to build meaningful shadow profiles, i.e. infer personal information on non-users of the platform that can be valuable for advertising?

Our empirical analysis relies on the entire browsing history of about 5'000 representative U.S. internet users, combined with information on whether the visited websites interacted with Facebook servers through engagement buttons. This allows us to measure Facebook's ability to track users across the web. Importantly, we can distinguish between Facebook users and user that choose not to visit the Facebook platform. After documenting the extent of surveillance Facebook can engage in with respect to anyone's web browsing activities, 
we show that the data Facebook can collect through its engagement buttons is useful to infer personal information about non-users of Facebook. Users of Facebook reveal personal information, such as birthday, gender, education and occupation to the platform. Because our data allow us to observe individuals' demographic information, we can mimic what Facebook could do and train a machine learning model to predict demographic outcomes -- such as age, gender, education, and income -- based on the web browsing activity of Facebook users. Given that the web browsing activity of users and non-users of Facebook overlaps, we can then use this model to predict demographic outcomes of non-users of Facebook.

Our study has important implications for ongoing debates around privacy regulation and competition policy. For individual users, privacy concerns could be reduced if tracking by Facebook outside of its own platform could easily be avoided by quitting Facebook or by visiting websites without engagement buttons.
While recent privacy regulations such as the EU's General Data Protection Regulation (GDPR), the California Consumer Privacy Act (CCPA) and industry changes such as Apple's blocking of third-party cookies have made third party tracking more difficult, the reality of tracking remains much the same. Legal ascertainment of consumer consent has become more robust over time but consumers effectively still do not know they consented or have little choice.\footnote{The Facebook antitrust case at the German federal cartel office was focused on consent and consumer choice, see \href{https://www.reuters.com/article/us-facebook-germany-idUSKCN1PW0SW}{https://www.reuters.com/article/us-facebook-germany-idUSKCN1PW0SW}, accessed 24 September 2021.} 

The EU's planned Digital Markets Act (DMA) will introduce new regulations on the use of data by the largest platforms, building on the notions of user consent and purpose limitation introduced by GDPR. However, worries remain that gaps in enforcement of GDPR may lead to similar ambiguities with the DMA.\footnote{See \href{https://techcrunch.com/2022/04/22/google-facebook-apple-eu-lobbying-report/}{https://techcrunch.com/2022/04/22/google-facebook-apple-eu-lobbying-report/}, accessed 20 May 2022.} The extent of tracking we document is indicative of the potential privacy risk due to shadow profiling for advertising purposes irrespective of technological or legal barriers. Given that large platforms observe more than just the basic demographics we can analyze here, our aim was to quantify a lower bound to the possibilities of shadow profiling online.

\section{Empirical analysis}

\subsection{Shadow Profiling}

Online platforms sell consumer profiles based on demographic information and interests to advertisers \cite{neumann2019frontiers}. Facebook can rely on precise individual targeting on its own platform by exploiting information on its users' demographic characteristics, their granular browsing behavior, their likes, and their social interactions. Previous research has shown, for example, that Facebook users' income can be predicted by their likes and status updates \cite{matz2019predicting}. 

Since 2016, Facebook has also sold targeted advertisement slots based on consumer profiles outside of its own platform. However, targeting consumers outside of its platform is more difficult as Facebook has much less data available for non-users of the platform. For instance, Facebook users typically share their demographic information -- such as age, gender, or occupation -- with the platform, but Facebook cannot directly obtain such information from non-users. Instead, the platform can rely on the fact that the off-Facebook browsing behavior of its users and non-users partially overlaps to predict the demographic characteristics of non-users. Such predictions can be performed even if users of Facebook never shared any information about non-users of Facebook, such as the sharing of contact lists documented in \cite{garcia2017shadowprof}, and even if non-users refuse to directly reveal their demographic information with the platform. This is the essence of shadow profiling, which allows large platforms such as Facebook to learn about any Internet user's personal characteristics without their knowledge or consent. 

In economic terms, Facebook users effectively impose an information or data externality \cite{maccarthy2010new,choi2019privacy,acemoglutoo,bergemannetal2022socialdata} on non-users through the revelation of their demographic information to the platform. We provide a quantification of Facebook's ability to predict non-users' demographic characteristics using domain-level web browsing data.

\subsection{Data}

We have access to individual-level desktop browsing data of a representative sample of the U.S. population via the market research firm Nielsen. Participants are incentivized to install a software that records all web browsing activity and fill in a survey of basic demographics, such as gender, employment, age, education, and income. We classify individuals who refused to answer a question about their income as privacy sensitive \cite{goldfarbtucker2012privsensitive}. For each user, we observe the web addresses (URLs) of all websites visited in 2016, as well as the time spent on each URL. We define Facebook users as those who visit Facebook at least once in that year. Nielsen further provides a categorization of websites on which we base our analysis of privacy-sensitive websites and competitors to Facebook.

Historical information on websites' connections to Facebook comes from the HTTPArchive project, which periodically crawls a large number of websites to record data on their use of third-party technology (including cookies). We use information collected on June 1, 2016 and filter connections to Facebook servers. 

Clickstream data has been widely used in the literature. Prior research has established that data from Nielsen does indeed provide a good representation of the population with Internet access \cite{aguiar18isr}. It is possible that users that sign up for Nielsen's service have different privacy preferences than the general population. The share of participants that choose not to reveal their income in our data is 2\% compared to 15\% in data from ad-hoc web surveys used in the literature \cite{goldfarbtucker2012privsensitive}. Note, however, that this is only a concern in our setting if privacy preferences affect the scope and intensity of web browsing. Users typically cannot know whether a website has Facebook's engagement buttons before they visit the website. With these caveats, it remains to note that clickstream data collected with voluntary consent of users is the best available data source for our study.

Finally, both within our sample and in general, Internet users may explicitly opt out of web tracking. However, awareness of tracking and opting out is rare in practice and requires significant user sophistication \cite{melicher2016preferences,mathur2018characterizing,weinshel2019oh}.

\subsection{Estimating online tracking intensity}

Combining the two datasets, we can compute the share of websites a user visits that make requests to Facebook servers, i.e. the share that can be tracked by Facebook. Specifically, we jointly observe, for each user, the share of domains and browsing time tracked $Y$, the sum of domains visited and browsing time spent per user as two measures of browsing intensity $W$, discretized personal characteristics $X$, Facebook user status $D$, and visited website categories $K$. We provide estimates of $E[Y|W=w_p,D]$, $E[Y|W=\bar{w},D,X]$, and $E[Y|W=\bar{w},D,K]$, where $w_p$ denotes browsing intensity by percentiles $p \in \{0.2,0.4,0.6,0.8,1.0\}$ and $\bar{w}$ denotes mean browsing intensity.

One potential concern is that some of the websites visited by individual users in the clickstream data are not observed in the HTTPArchive project. In other words, we are unable to obtain websites' connections to Facebook for a subset of domains in our clickstream data. In order to quantify the extent of this selection problem, we estimate Manski bounds \cite{manski1995}. More specifically, of the 491'841 website domains visited by 4,989 users in our data, 109'512 are observed in HTTPArchive, $(z=1)$. Domains not observed in HTTPArchive, $(z=0)$, account for $P(z=0)=0.25$ of unique domains visited and $P(z=0)=0.11$ of total browsing time. Decomposing $E[Y] = E[Y|z=1] P(z=1) + E[Y|z=0] P(z=0)$, we note $E[Y|z=0]$ cannot be estimated. However, we know that $0 \leq E[Y|z=0] \leq 1$ and so can estimate bounds of $E[Y]$: $E[Y|z=1] P(z=1) \leq E[Y] \leq E[Y|z=1] P(z=1) + P(z=0)$.

\subsection{Prediction method}

One approach to form consumer audiences for targeted advertising is the prediction of users' membership in demographic groups \cite{neumann2019frontiers}. We construct classification models to predict individual users' membership in 10 binary demographic sub-groups: four age ranges (0-17, 18-24, 25-34, 35-45), the presence of children in household, female gender, high education level, high income, residence in political swing state, and unemployed status. We drop 136 users who visit no domains tracked by Facebook, accounting for roughly 10\% of non-users of Facebook. We observe demographic group membership for 3,747 Facebook users accounting for 17.67 million clicks in the sample year and for 1,106 non-users of Facebook accounting for 0.5 million clicks in the sample year. To render the clickstream data into a form amenable to machine learning, we define each domain in the clickstream data as a feature. Each feature takes on a continuous value measuring the number of clicks on the domain throughout the sample year. These data are structured, highly sparse, and the conditional expectation function of the predicted class contains potentially rich interactions between large numbers of domains.

To tackle this prediction task, we use the Extreme Gradient Boosting algorithm (XGBoost) with the binary logistic loss function \cite{hastie2009elements,chen2016xgboost}. As Facebook can only observe demographics its users directly share with the platform, we tune the prediction algorithm only using the Facebook user sample. Given the number of prediction tasks, we automatize the step of hyperparameter tuning for the XGBoost algorithm. We use a Bayesian optimization method to tune hyperparameters using the hyperopt package \cite{bergstra2013making} based on 5-fold cross validation. Given hyperparameters, we train the algorithm on 100 bootstrap splits of the Facebook user sample into a training partition and an evaluation partition. To obtain comparable out-of-sample predictions, we choose the split such that the evaluation partitions are of similar size as the sample of non-users of Facebook. 

For the Facebook sample, we report the bootstrap aggregated mean and standard errors for two prediction quality measures. For tuning and training, the target prediction measure is the area under the receiver operation characteristic (ROC) curve (AUC). The AUC, bounded by [0.5,1], quantifies the location of the ROC curve, which represents all achievable trade-offs between false positive rates and true positive rates by a given prediction technology. As a second prediction quality measure, we report balanced accuracy to take into account that some demographic groups are imbalanced such as high income or children in household. Because we do not use the sample of non-users of Facebook to train the algorithm, we compute the confidence intervals for the prediction quality measures using 100 bootstrap samples from that sample.

\section{Results}

\subsection{What is the Extent of Facebook's Online Tracking?}

We start by showing our results regarding Facebook's ability to track individuals' online browsing across different types of user demographics and domains.

\subsubsection{Tracking across Different Users}

\begin{figure}[h!]
\centering
\caption{Extent of Facebook web tracking by user type}
    \includegraphics[width=10cm]{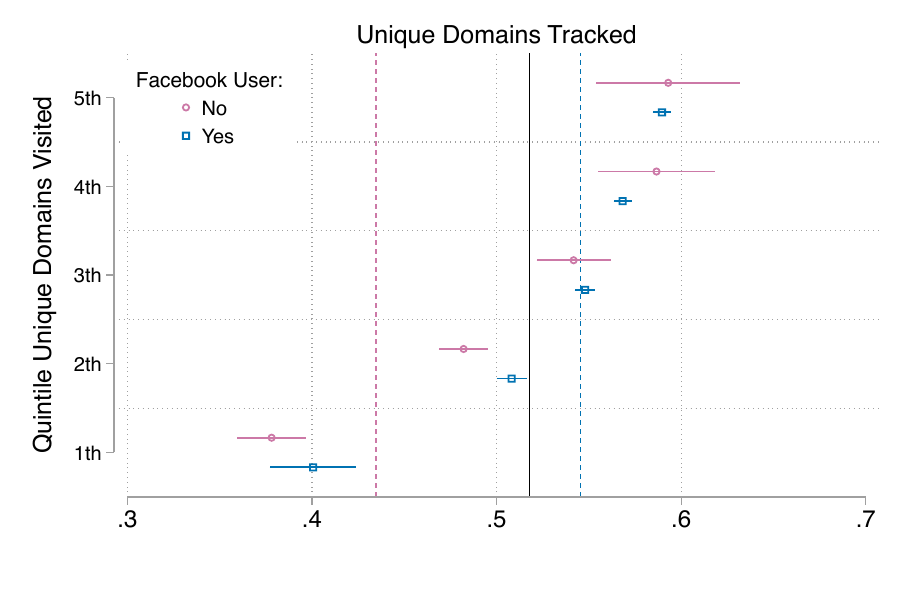}\\[1em]
    \includegraphics[width=10cm]{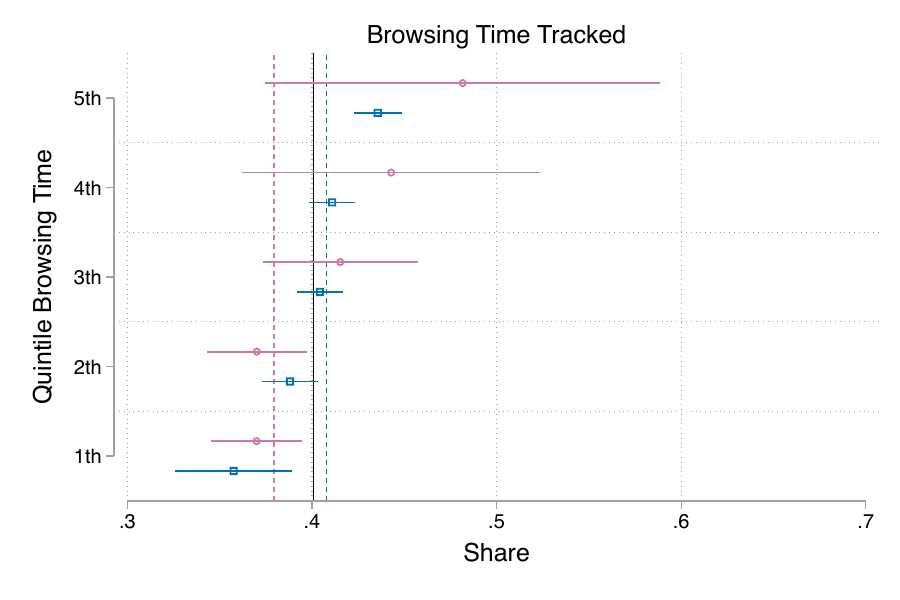}
\label{fig:results1}
\begin{figurenotes}
The Figure shows mean shares of the number of unique visited websites and browsing time tracked by quintiles of each browsing intensity measure, for Facebook users and non-Facebook users. We define a person as Facebook if the Facebook web domain was visited at least once in the sample year. In the first panel, the solid line shows the mean share of websites tracked, 52\%, and dashed vertical lines show the mean shares of websites tracked for Facebook users and non-Facebook users, 55\% and 44\%. In the second panel, the solid line shows the mean share of browsing time tracked, 40\%, and dashed vertical lines show the mean shares of browsing time tracked for Facebook users and non-Facebook users, 41\% and 38\%.
\end{figurenotes}
\end{figure}

\begin{figure}[htbp]
\centering
\caption{Extent of Facebook web tracking by user type}
    \includegraphics[width=10cm]{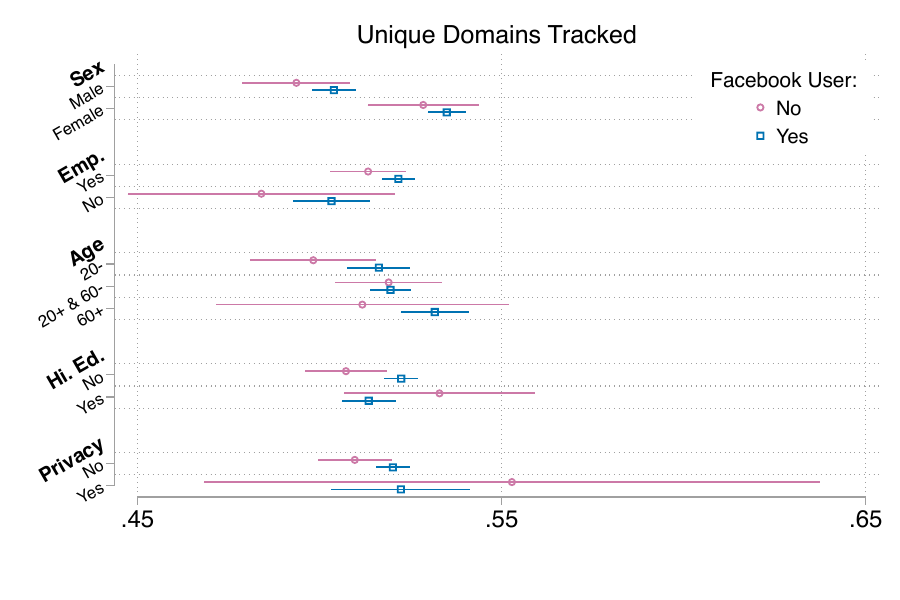}\\[1em]
    \includegraphics[width=10cm]{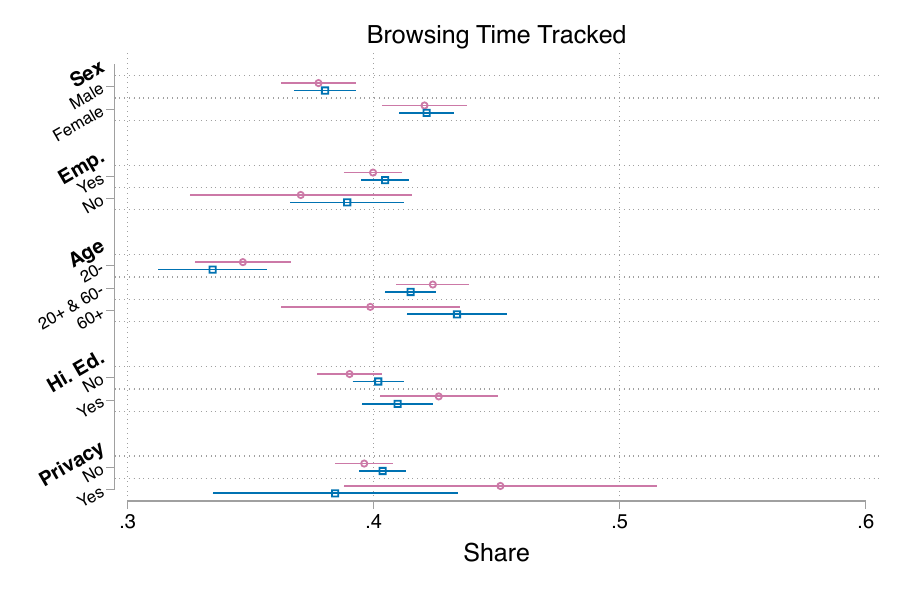}
\label{fig:results2}
\begin{figurenotes}
The Figure shows mean shares of visited websites and browsing time tracked, conditional on the respective demographic characteristic and mean browsing intensity, for Facebook users and non-Facebook users. We define a person as Facebook if the Facebook web domain was visited at least once in the sample year.
\end{figurenotes}
\end{figure}

Figures \ref{fig:results1} and \ref{fig:results2} show the extent to which Facebook can track its users and non-users, and how this depends on overall browsing intensity and demographics. The top panel in Figure \ref{fig:results1} focuses on the total number of websites that individuals visit during the sample period, excluding Facebook. As shown by the two vertical dashed lines, 55\% of websites visited by Facebook users and 44\% of websites visited by non-Facebook users are tracked, on average, by Facebook. For a given individual, the share of their visited websites that can be tracked by Facebook increases with their total online activity, measured in quintiles of the total number of websites visited (including Facebook). The extent of Facebook's potential tracking -- conditional on online activity -- is similar for users and non-users of Facebook. Hence, differences in the extent of overall tracking, denoted by the dashed vertical lines, are due to Facebook users' higher online activity.

The number of visited websites may not fully capture the intensity with which individuals browse the internet. The lower panel of Figure \ref{fig:results1} reports mean shares of total browsing time that Facebook can track over the sample period. We observe smaller differences in the average share of online activity that can be tracked for users and non-users of Facebook, depicted as vertical dashed lines at 41\% and 38\%. This suggests that Facebook is similarly well connected to websites on which users and non-users of Facebook spend a similar amount of time. Perhaps surprisingly, the share of an individual's online browsing time that can be tracked by Facebook increases only marginally with online activity.

For website domains visited, the estimated Manski bounds are $[0.38,0.64]$ and for browsing time $[0.35,0.47]$. Absent any assumption on the selection of domains, these bounds remain informative and close to our point estimates. We therefore focus on point estimates when reporting all further results.

Figure \ref{fig:results2} shows the mean share of websites and browsing time that Facebook can track by demographic groups. The top panel shows some heterogeneity regarding the extent to which certain demographic groups are being tracked. For instance, females tend to be tracked more than male. We see little difference in tracking between individuals who are privacy sensitive and those who are not. We do not find significant differences in the share of domains tracked between users and non-users of Facebook for any demographic group. As these shares are conditional on mean browsing intensity, any differences would be ascribed to the types of websites visited by users, in terms of Facebook coverage, in their respective demographic group. Hence, Facebook benefits from engagement buttons being placed on a diverse set of websites. The lower panel further shows that, within demographic groups, there are no significant differences in the share of browsing time tracked for users and non-users of Facebook.\footnote{In order to check the robustness of the results to the definition of a Facebook user, we conduct similar estimations by classifying an individual as a Facebook user if they visited the Facebook domain more than 10 times during the sample period. The results then show that differences in tracking between users and non-users of Facebook increases only slightly.}

We have so far defined Facebook users as individuals who visited the Facebook domain at least once during our sample period. This definition naturally has some caveats. First, because our sample only allows us to observe web browsing on a desktop machine, we are not able to distinguish Facebook users who only engage with the platform through an alternative device. The individuals we refer to as ``non-users'' of Facebook can therefore also be interpreted as users that have no interest in visiting Facebook on a desktop computer. Second, one may worry about our definition of a Facebook user being too strict when comparing the extent of tracking between users and non-users of Facebook. If the two groups differ systematically in the websites they spend time on and if Facebook users are more likely to spend time on websites tracked by Facebook, then mis-classifying some non-users as Facebook users would downward bias the difference in the extent of tracking between the two groups. The level of tracking for non-users of Facebook should be unaffected if they are a homogeneous group with respect to tracking intensity. Figures \ref{fig:results1_10fbclicks} and \ref{fig:results2_10fbclicks} in the Appendix show that results are similar when using an alternative definition, where users with more than ten clicks on the Facebook domain per year are considered Facebook users. The difference in tracking between users and non-users of Facebook increases only slightly.

\subsubsection{Tracking across Types of Websites}

Table \ref{tab:track_domains} shows the share of websites and browsing time that can be tracked by Facebook across website categories. The first two columns show that the overall share of websites that are connected to Facebook servers is significantly higher for users of Facebook relative to non-users. A closer look at specific website categories nevertheless reveals that this difference is mainly driven websites related to instant messaging services, where Facebook users are tracked to a much larger extent than non-Facebook users.\footnote{Note that these instant messaging services do not include services offered by Facebook.} For the remainder of the website categories, we observe small differences between users and non-users of Facebook. The shares of unique domains and browsing time tracked are smaller overall for website categories that likely include online platforms competing with Facebook. This observation is consistent with a reduced level of integration of Facebook's engagement buttons on competing services, in line with individuals' interest in privacy online. 
When focusing further on browsing time in the last two columns, the differences between users and non-users of Facebook are much less pronounced. This holds overall and in all website categories except instant messaging. It is therefore apparent that there are differences in Facebook's overall ability to track visits and time spent across website categories. It is nevertheless worth noting that Facebook's engagement button technology can track attention online irrespective of domain categories, even those considered privacy-sensitive. This is true for users and non-users of Facebook alike, as shown by the small within-categories differences between these two groups.

\subsubsection{Tracking in Relation to Facebook Use}

\begin{figure}[htbp]
\centering
\caption{Browsing Time Composition}
    \includegraphics[width=\linewidth,trim=0 0 0 0 ]{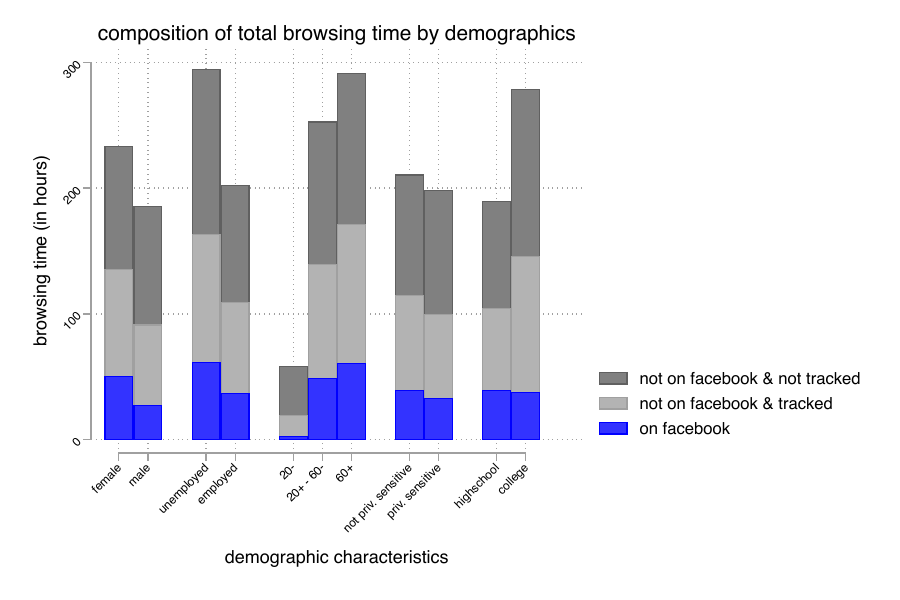}\\[1em]
\label{fig:results3}
\begin{figurenotes}
The Figure shows the average amount of browsing time an individual falling into a respective demographic spends on a) Facebook, b) websites with Facebook engagement buttons, and c) websites without Facebook engagement buttons.
\end{figurenotes}
\end{figure}

We further explore the distribution of browsing time across Facebook.com, websites that have Facebook's engagement buttons (``tracked''), and websites that do not have Facebook's engagement buttons (``not tracked''). Figure \ref{fig:results3} shows the average amount of time spent on either one of these three categories of website, by demographic. Across most demographic groups, the browsing time that is either directly or indirectly monitored by Facebook accounts for more than $50$ percent of the total browsing time.  

Across all demographics, the browsing time tracked by Facebook is larger than the actual time spent on Facebook's platform. While there is large variation in total surfing time across demographic groups, the relative shares of time spent on Facebook and on other websites tracked by Facebook are remarkably similar. Figure \ref{fig:results3} further highlights the scope of Facebook tracking when including the online activity it can directly monitor on its own platform. The network of engagement buttons appears to increase the amount of time tracked proportionally to the time spent on Facebook.com.

\subsubsection{Browsing Similarities across Users and Non-Users of Facebook}

We further ask whether the similarity in tracking ability between users and non-user of Facebook is driven by similarity in browsing behavior. Are users and non-users tracked similarly because they visit similar websites, i.e. because they have similar browsing profiles? 

We rely on the cosine similarity to construct a measure of browsing similarity across users and non-users of Facebook. The cosine similarity is based on the click-weighted website visits of each individual in the sample, which can be represented by a high-dimensional vector that records the number of times the individual clicked on any given website.\footnote{Consistent with the previous analysis, we exclude the Facebook domain when computing the cosine similarity.} Based on this number of click-weighted website visits, we compute the cosine-similarity between each pair of individuals, both within and between users and non-users of Facebook. This results in a measure ranging between zero and one with higher values indicating higher similarities.

Table \ref{tab:similarity_cos} reports the cosine similarities within and between users and non-users of Facebook. The average pairwise similarity over all users is $0.15$. The average pairwise similarity is highest within the group of Facebook users ($0.18$), above the overall average, and lowest within the group of non-users ($0.08$), below the overall average. The observation that similarity within non-users of Facebook is smaller than the overall average similarity suggests that non-users of Facebook display the most dispersed browsing behavior. Interestingly a non-user is, on average, more similar to a Facebook user than to another non-user ($0.12$). Despite this, the observation that similarities are larger within Facebook users than between Facebook users and non-users reveals differences in browsing behavior between the typical Facebook user and the typical non-user. We conclude that users and non-users of Facebook have overlapping but quite distinct browsing behaviors. It is challenging to assess, based on these descriptive observations alone, to what extent a platform such as Facebook can make use of information reflected by consumer browsing behavior.

\subsection{Shadow Profiles: Predicting Demographic Information}

To quantify the extent to which Facebook may successfully engage in the profiling of non-users of the platform, we use our sample data and train a machine learning algorithm to predict 10 demographic characteristics of Facebook users based on their off-platform browsing behavior. Table \ref{tab:shprofile_predict} shows that prediction quality is high for the sample of Facebook users, with AUC values exceeding 0.65 for all for all binary demographic characteristics, except for classifying individuals' residence in a swing state or their unemployment status. The highest prediction quality is achieved for predicting low user age and the presence of children in the household. 

We can compare our prediction accuracy results with existing studies analyzing the accuracy of third-party consumer profiling. \cite{neumann2019frontiers} use field studies to evaluate the accuracy of data brokers' consumer profiling on various demographic and audience-interest attributes. In the specific case of gender, they find that prediction accuracy varies greatly across 14 data brokers, with levels ranging from 26\% to 63\%. They report an average accuracy of 43\%, which is below the accuracy of random draws, 50\%, that would achieved without consumer profiling. Predictions based on age tiers identical to ours reveal accuracy levels ranging from 7\% to 44\%, which imply on average 42\% accuracy improvement compared to no targeting. We report balanced accuracy, for which a value of 50\% is achieved without targeting. For the same age brackets as in \cite{neumann2019frontiers}, we find improvements over no information lie between 24\% and 60\%, on average 40\%. Children in households can be predicted with 40\% higher accuracy than with no information, and female gender with 16\% higher accuracy.

For the sample of non-users of Facebook, we find lower prediction quality than for the sample of Facebook users, with differences that are statistically significant at the five percent level. However, in particular the demographic characteristics low age, presence of children in the household, and high education status can be predicted well, providing meaningful inference of personal characteristics. In particular, non-FB users can be classified into different age bins with accuracy that are up to 30\% higher than with no information. Children in household can be predicted with 14\% higher accuracy than without information, and female gender with 6\% higher accuracy. For an alternative definition of FB user status, we find a slightly higher prediction accuracy throughout as reported in Table \ref{tab:shprofile_predict_FB_l10} in the Appendix.

We finally highlight that our data allow us to quantify a conservative estimate of Facebook's ability to build user profiles based on broad demographic characteristics. Facebook users share further personal and preference-related information with the platform which would allow building richer individual consumer profiles for user outside of the Facebook platform.

\section{Conclusion}

Our results document the extent to which a large online platform such as Facebook can build shadow profiles based on users' web browsing behavior outside of its core platform. We document an indiscriminate ability to collect user data, independent of user characteristics such as demographics and whether they actively use Facebook's social networking platform.

Information that lets firms infer consumer preferences can be used to extract rents, for example through consumer steering, targeted advertising, or price discrimination. When large platforms have access to extensive consumer data, this may lead to further concentration in the online advertising industry in which network effects play a crucial role. Resulting higher prices for advertisers increase their cost which in turn result in higher product prices. By documenting that a large platform like Facebook is able to collect data on consumers that do not use their platform, our results suggest further potential harms to consumers. Even users that do not receive utility from using Facebook are subject to the externalities that Facebook's data collection may impose on them.

One important limitation of this study is that we quantify the possibility of shadow profiling based on a sample which contains significantly less data than what Facebook collects. Facebook can see richer personal information and user's consumption decisions online, which it may attempt to predict for more sophisticated consumer profiles and consumer steering. While we focus on a sample of one year, Facebook can see longer browsing histories. In addition, Facebook may be able to exploit context-dependent browsing paths, while we restrict our analysis to the level of web domains. Even with our limited data, however, we find important data externalities, or spill overs, from Facebook users to non-users of Facebook. A further limitation is that we focus on the task of predicting personal information. Ultimately, these predictions may be used to steer users' online activities so there is important room for research on how valuable shadow profiles can be in terms of impacting individuals' access to information or consumers' decisions to purchase products online.

Regulatory efforts around the globe have tried to reduce the type of de-facto tracking that we highlight above. Examples include the European Union's (EU) ePrivacy Directive from 2003 that regulated opt-in to data collection via cookies as well as the EU GDPR in 2018. Among other things, the latter mandated informed consent to data collection and introduced harsh fines. Similar legislation was introduced in the US with the CCPA that became effective in 2020. Leading actors in the industry such as Apple and Google are recently orchestrating a move away from third-party cookies for tracking. However, cookies are or will be replaced by new generations of tracking technologies, such as device fingerprinting, software development kits, and in-app browsing. The ability of Facebook to create shadow profiles is therefore not limited to Facebook's current technology of cookies and engagement buttons, and perhaps more importantly not limited to Facebook. Google places third-party scripts on 80\% of the top 1000 websites \cite{accc2021} and has shifted to using its Chrome browser to track online activities. Hence, despite regulatory efforts and technological change, the potential for indiscriminate large-scale tracking, such as using third-party scripts as we document, is likely to remain.

\newpage
\bibliographystyle{aea}
\bibliography{facebook}

\newpage
\section*{Figures and Tables}


\clearpage
\begin{table}[htbp] \centering
\begin{threeparttable}
\caption{Shadow profiling by types of websites}
\label{tab:track_domains}
\begin{tabular}{@{} l @{\extracolsep{1mm}}  c  @{\extracolsep{3mm}} c @{\extracolsep{5mm}} c  @{\extracolsep{3mm}} c @{}}
\toprule
                    & \multicolumn{2}{c}{Unique domains}    & \multicolumn{2}{c}{Browsing time} \\
                    \cmidrule{2-3}\cmidrule{4-5}
Website category     & non-FB user & FB user                 & non-FB user & FB user       \\ 
\midrule\\[-0.25cm]
All$^{\star  \dag}$                         & 0.44 & 0.55 & 0.38 & 0.41 \\[0.1cm]
\emph{Privacy-sensitive} &&&&\\
Adult$^{\star}$                            & 0.09 & 0.13 & 0.09 & 0.12 \\
Career Development              & 0.42 & 0.45 & 0.37 & 0.42 \\
Dating                          & 0.54 & 0.53 & 0.55 & 0.55 \\
Finance/Insurance/Investment    & 0.45 & 0.46 & 0.45 & 0.45 \\
Gambling                        & 0.66 & 0.62 & 0.68 & 0.67 \\
Government                      & 0.15 & 0.16 & 0.15 & 0.15 \\
Health, Fitness \& Nutrition    & 0.63 & 0.64 & 0.63 & 0.64 \\
Real Estate/Apartments          & 0.74 & 0.75 & 0.78 & 0.79 \\
[0.1cm]
\emph{Competing platforms} &&&&\\
E-mail$^{\star}$                          & 0.01 & 0.04 & 0.01 & 0.01 \\
Instant Messaging$^{\star  \dag}$                & 0.08 & 0.23 & 0.05 & 0.19 \\
Member Communities              & 0.58 & 0.60 & 0.58 & 0.58 \\
Search$^{\star  \dag}$                           & 0.21 & 0.29 & 0.22 & 0.29 \\
\bottomrule
\end{tabular}
\begin{tablenotes}
\item \emph{Notes}: The table reports mean user-level shares of unique domains and browsing time tracked by Facebook (FB).
$\star$: Difference in unique domains between user groups are statistically significant at the five percent level.
$\dag$: Difference in browsing time between user groups are statistically significant at the five percent level.
\end{tablenotes}
\end{threeparttable}
\end{table}

\clearpage

\begin{table}[htbp] \centering
\begin{threeparttable}
\caption{Cosine Similarities}
\label{tab:similarity_cos}
	\begin{tabular}{@{} l @{\extracolsep{1mm}}  l  @{\extracolsep{5mm}} c @{\extracolsep{2mm}} c  @{\extracolsep{7.5mm}} c @{\extracolsep{2mm}} c @{\extracolsep{2mm}} c @{}} 
			\toprule
			\multicolumn{1}{c}{} &  & \multicolumn{2}{c}{non-FB user}    & &  \multicolumn{2}{c}{FB user}   \\[0.1em]
			\midrule
			 & \multirow{2}*{non-FB user}   & $0.0824$              & \multirow{2}*{$N = 0.7$M} & &  &                                  \\
			 &                              & $[0.0821, 0.0828]$    &                           & &  &                                  \\
			 \midrule
			 & \multirow{2}*{FB user}       & $0.1198$              & \multirow{2}*{$N = 4.6$M} & & $0.1768$ & \multirow{2}*{$N = 7$M}  \\
			 &                              & $[0.1196, 0.1199]$    &                           & & $[0.1767, 0.1770]$ &                \\
			 \bottomrule
		\end{tabular}
\begin{tablenotes}
\item \emph{Notes}: The table reports the average cosine similarity between all consumer pairs belonging to the respective groups. The square brackets report the corresponding $95\%$ confidence intervals. The number of consumer pairs used for calculation is reported in millions. The average cosine similarity between any consumer pair in the data is $0.1495$, the corresponding confidence interval is $[0.1495, 0.1497]$.  
\end{tablenotes}
\end{threeparttable}
\end{table}

\begin{table}[h!]
\centering
\begin{threeparttable}
\caption{Predicting demographics for users and non-users of FB}
\label{tab:shprofile_predict}
\begin{tabular}{ @{} l @{\extracolsep{8mm}} c @{\extracolsep{4mm}} c @{\extracolsep{4mm}} c @{\extracolsep{10mm}} c @{\extracolsep{4mm}} c @{\extracolsep{4mm}} c @{} }
\toprule
& \multicolumn{3}{c}{AUC} & \multicolumn{3}{c}{Balanced accuracy} \\
\cline{2-4} \cline{5-7}
                        & FB & non-FB & $\Delta$& FB & non-FB & $\Delta$\\
\midrule
Age 0-17		        & 0.88 & 0.77 & 0.12	    & 0.80 & 0.65 &0.15     \\[-0.3em]&{\scriptsize(0.011)}&{\scriptsize(0.014)}&{\scriptsize(0.018)}& {\scriptsize(0.019)}&{\scriptsize(0.007)}&{\scriptsize(0.020)}\\
Age 18-24		        & 0.80 & 0.62 & 0.18	    & 0.71 & 0.55 &0.15     \\[-0.3em]&{\scriptsize(0.021)}&{\scriptsize(0.029)}&{\scriptsize(0.036)}& {\scriptsize(0.019)}&{\scriptsize(0.013)}&{\scriptsize(0.023)}\\
Age 25-34		        & 0.73 & 0.59 & 0.14	    & 0.66 & 0.56 &0.09	    \\[-0.3em]&{\scriptsize(0.015)}&{\scriptsize(0.026)}&{\scriptsize(0.030)}& {\scriptsize(0.020)}&{\scriptsize(0.017)}&{\scriptsize(0.027)}\\
Age 35-45		        & 0.66 & 0.55 & 0.11        & 0.62 & 0.50 &0.12     \\[-0.3em]&{\scriptsize(0.021)}&{\scriptsize(0.019)}&{\scriptsize(0.028)}& {\scriptsize(0.016)}&{\scriptsize(0.007)}&{\scriptsize(0.018)}\\
Children in HH	        & 0.78 & 0.70 & 0.08	    & 0.70 & 0.57 &0.13 	\\[-0.3em]&{\scriptsize(0.017)}&{\scriptsize(0.016)}&{\scriptsize(0.023)}& {\scriptsize(0.013)}&{\scriptsize(0.006)}&{\scriptsize(0.015)}\\
Female                  & 0.76 & 0.62 & 0.15	    & 0.58 & 0.53 &0.05     \\[-0.3em]&{\scriptsize(0.014)}&{\scriptsize(0.014)}&{\scriptsize(0.020)}& {\scriptsize(0.012)}&{\scriptsize(0.006)}&{\scriptsize(0.013)}\\
Higher education        & 0.74 & 0.68 & \emph{0.06}	& 0.68 & 0.57 &0.11	    \\[-0.3em]&{\scriptsize(0.021)}&{\scriptsize(0.024)}&{\scriptsize(0.032)}& {\scriptsize(0.013)}&{\scriptsize(0.012)}&{\scriptsize(0.018)}\\
High income	            & 0.68 & 0.61 & 0.07	    & 0.57 & 0.50 &0.07     \\[-0.3em]&{\scriptsize(0.013)}&{\scriptsize(0.014)}&{\scriptsize(0.020)}& {\scriptsize(0.009)}&{\scriptsize(0.002)}&{\scriptsize(0.009)}\\
In swing state	        & 0.64 & 0.54 & 0.10	    & 0.59 & 0.51 &0.07	    \\[-0.3em]&{\scriptsize(0.024)}&{\scriptsize(0.018)}&{\scriptsize(0.030)}& {\scriptsize(0.016)}&{\scriptsize(0.006)}&{\scriptsize(0.017)}\\
Unemployed              & 0.60 & 0.60 &\emph{-0.01}	& 0.55 & 0.50 &0.05     \\[-0.3em]&{\scriptsize(0.028)}&{\scriptsize(0.026)}&{\scriptsize(0.038)}& {\scriptsize(0.019)}&{\scriptsize(0.016)}&{\scriptsize(0.025)}\\
\midrule
Number of users 	    &3,747 &1,106 &         &3,747 &1,106 &       \\
Total clicks (in 1,000) &17,671&499   &         &17,671&499   &       \\
\bottomrule
\end{tabular}
\begin{tablenotes}[flushleft]
\item Notes: This table reports AUC and balanced accuracy as measures of prediction quality using the XGBoost algorithm for the sample of FB users and non-users. An individual is classified as a FB user if the FB domain was visited at least once in the sample year. Standard errors are based on 100 bootstrap samples. Differences in italic are not statistically significantly different from zero at the 5\% level.
\end{tablenotes}
\end{threeparttable}
\end{table}

\clearpage
\section*{Appendix} 

\setcounter{table}{0}
\setcounter{figure}{0}
\renewcommand{\thetable}{A.\arabic{table}}
\renewcommand{\thefigure}{A.\arabic{figure}}


\begin{figure}[h!]
\centering
\caption{Extent of Facebook web tracking by user type (alternative definition)}
    \includegraphics[width=10cm]{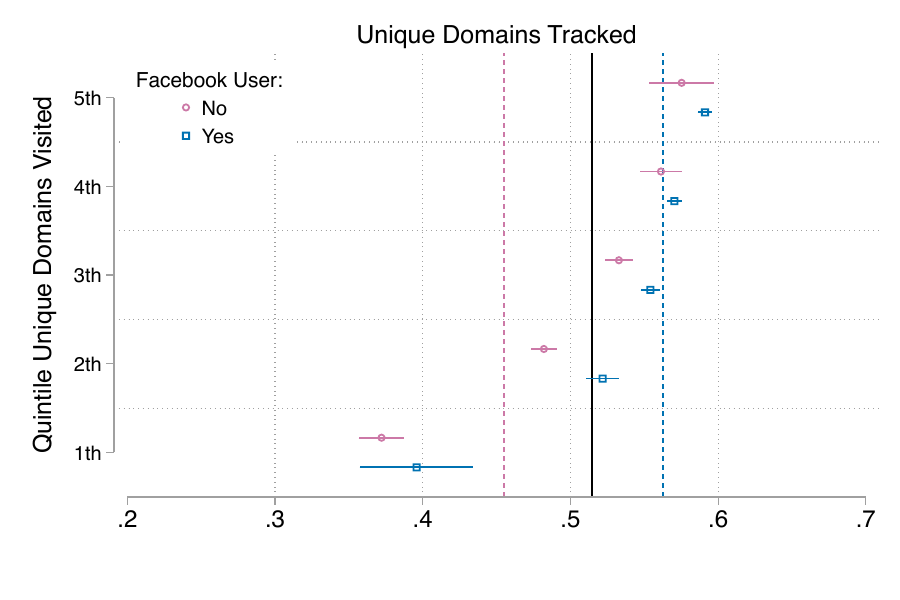}\\[1em]
    \includegraphics[width=10cm]{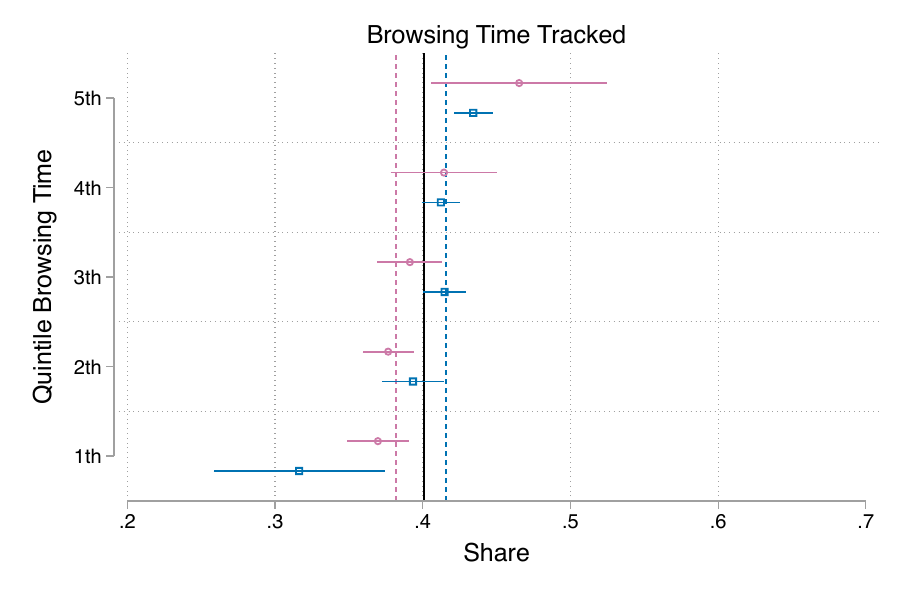}
\label{fig:results1_10fbclicks}
\begin{figurenotes}
The Figure shows mean shares of the number of unique visited websites and browsing time tracked by quintiles of each browsing intensity measure, for Facebook users and non-Facebook users. In this Figure, we define a person as Facebook user if the Facebook web domain was visited more than 10 times in the sample year. In the first panel, the solid line shows the mean share of websites tracked, 52\%, and dashed vertical lines show the mean shares of websites tracked for Facebook users and non-Facebook users, 55\% and 44\%. In the second panel, the solid line shows the mean share of browsing time tracked, 40\%, and dashed vertical lines show the mean shares of browsing time tracked for Facebook users and non-Facebook users, 41\% and 38\%.
\end{figurenotes}
\end{figure}

\begin{figure}[htbp]
\centering
\caption{Extent of Facebook web tracking by user type (alternative definition)}
    \includegraphics[width=10cm]{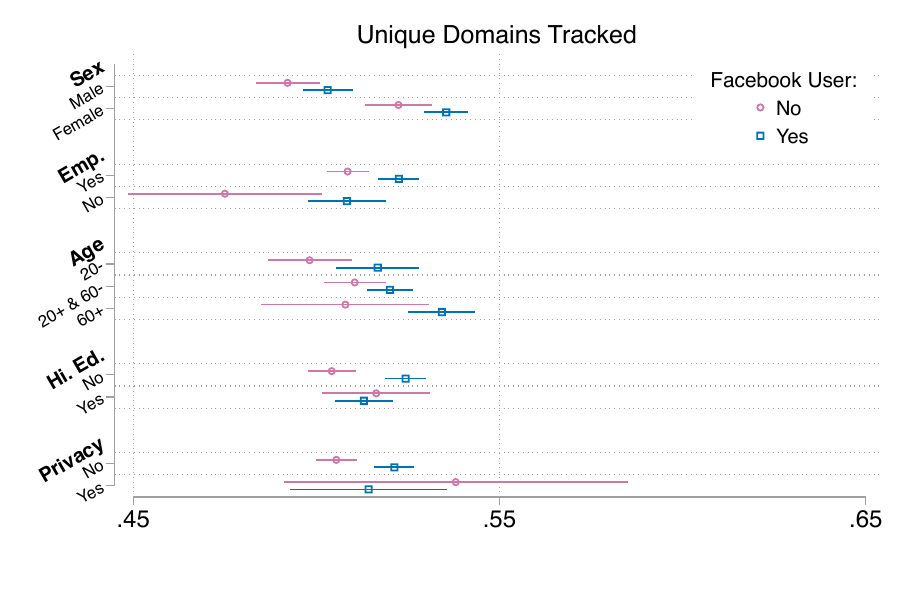}\\[1em]
    \includegraphics[width=10cm]{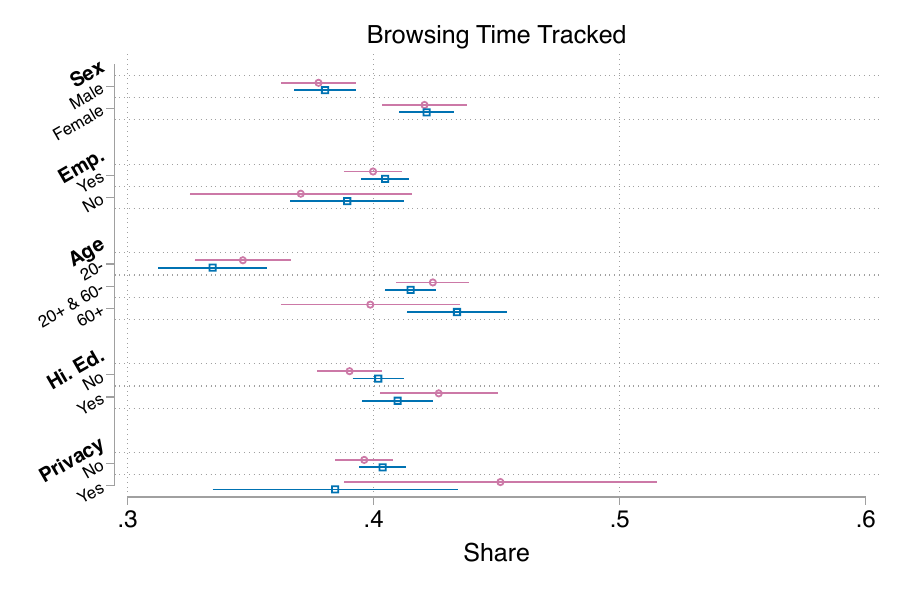}
\label{fig:results2_10fbclicks}
\begin{figurenotes}
The Figure shows mean shares of visited websites and browsing time tracked, conditional on the respective demographic characteristic and mean browsing intensity, for Facebook users and non-Facebook users. In this Figure, we define a person as Facebook user if the Facebook web domain was visited more than 10 times in the sample year.
\end{figurenotes}
\end{figure}

\begin{table}[h!]
\centering
\begin{threeparttable}
\caption{Predicting demographics for users and non-users of FB (alternative definition)}
\label{tab:shprofile_predict_FB_l10}
\begin{tabular}{ @{} l @{\extracolsep{8mm}} c @{\extracolsep{4mm}} c @{\extracolsep{4mm}} c @{\extracolsep{10mm}} c @{\extracolsep{4mm}} c @{\extracolsep{4mm}} c @{} }
\toprule
& \multicolumn{3}{c}{AUC} & \multicolumn{3}{c}{Balanced accuracy} \\
\cline{2-4} \cline{5-7}
                        & FB & non-FB & $\Delta$& FB & non-FB & $\Delta$\\
\midrule
Age 0-17		        & 0.86 &0.75 & 0.11	    	& 0.76 & 0.64 & 0.12	    \\[-0.3em]&{\scriptsize(0.019)}&{\scriptsize(0.012)}&{\scriptsize(0.022)}& {\scriptsize(0.019)}&{\scriptsize(0.007)}&{\scriptsize(0.020)}\\
Age 18-24		        & 0.80 &0.67 & 0.13	    	& 0.69 & 0.62 & 0.07	    \\[-0.3em]&{\scriptsize(0.022)}&{\scriptsize(0.019)}&{\scriptsize(0.029)}& {\scriptsize(0.019)}&{\scriptsize(0.013)}&{\scriptsize(0.023)}\\
Age 25-34		        & 0.70 &0.65 & \emph{0.05}	& 0.64 & 0.60 & \emph{0.04}	\\[-0.3em]&{\scriptsize(0.016)}&{\scriptsize(0.028)}&{\scriptsize(0.033)}& {\scriptsize(0.020)}&{\scriptsize(0.017)}&{\scriptsize(0.027)}\\
Age 35-45		        & 0.64 &0.57 & 0.07         & 0.60 & 0.52 & 0.07        \\[-0.3em]&{\scriptsize(0.020)}&{\scriptsize(0.013)}&{\scriptsize(0.024)}& {\scriptsize(0.016)}&{\scriptsize(0.007)}&{\scriptsize(0.018)}\\
Children in HH	        & 0.74 &0.74 &\emph{-0.002}	& 0.67 & 0.59 & 0.09	    \\[-0.3em]&{\scriptsize(0.018)}&{\scriptsize(0.009)}&{\scriptsize(0.020)}& {\scriptsize(0.013)}&{\scriptsize(0.006)}&{\scriptsize(0.015)}\\
Female                  & 0.76 &0.65 & 0.11        	& 0.53 & 0.55 &\emph{-0.02} \\[-0.3em]&{\scriptsize(0.011)}&{\scriptsize(0.007)}&{\scriptsize(0.013)}& {\scriptsize(0.012)}&{\scriptsize(0.006)}&{\scriptsize(0.013)}\\
Higher education        & 0.72 &0.70 & \emph{0.01}	& 0.65 & 0.60 & 0.05	    \\[-0.3em]&{\scriptsize(0.018)}&{\scriptsize(0.015)}&{\scriptsize(0.023)}& {\scriptsize(0.013)}&{\scriptsize(0.012)}&{\scriptsize(0.018)}\\
High income	            & 0.66 &0.61 & 0.05	    	& 0.52 & 0.51 & 0.02	    \\[-0.3em]&{\scriptsize(0.009)}&{\scriptsize(0.003)}&{\scriptsize(0.009)}& {\scriptsize(0.009)}&{\scriptsize(0.002)}&{\scriptsize(0.009)}\\
In swing state	        & 0.55 &0.60 &\emph{-0.05}	& 0.54 & 0.54 &\emph{-0.00}	\\[-0.3em]&{\scriptsize(0.014)}&{\scriptsize(0.006)}&{\scriptsize(0.015)}& {\scriptsize(0.016)}&{\scriptsize(0.006)}&{\scriptsize(0.017)}\\
Unemployed              & 0.57 &0.50 & 0.07	    	& 0.54 & 0.52 & \emph{0.02}	\\[-0.3em]&{\scriptsize(0.024)}&{\scriptsize(0.000)}&{\scriptsize(0.024)}& {\scriptsize(0.019)}&{\scriptsize(0.016)}&{\scriptsize(0.025)}\\
\midrule
Number of users 	    &2,754  &2,099  &         &2,754  &2,099  &       \\
Total clicks (in 1,000) &16,375 &1,796  &         &16,375 &1,796  &       \\
\bottomrule
\end{tabular}
\begin{tablenotes}[flushleft]
\item Notes: This table reports AUC and balanced accuracy as measures of prediction quality using the XGBoost algorithm for the sample of FB users and non-users. An individual is classified as a FB user if the FB domain was visited more than 10 times in the sample year. Standard errors are based on 100 bootstrap samples. Differences in italic are not statistically significantly different from zero at the 5\% level.
\end{tablenotes}
\end{threeparttable}
\end{table}

\end{document}